\documentclass[aps,graphicx,amsmath,amssymb,superscriptaddress,reprint]{revtex4-1}

\usepackage[dvipdfmx]{graphicx}
\usepackage{dcolumn}
\usepackage{bm}
\usepackage[colorlinks,linkcolor=blue,citecolor=blue]{hyperref}

\begin{document}

\title{Magnetic excitations in the noncentrosymmetric magnet Sr$_2$MnSi$_2$O$_7$}

\author{Masahiro Kawamata}
\affiliation{Department of Physics, Tohoku University, Sendai, Miyagi 980-8578, Japan}
\author{Xiaoqi Pang}
\affiliation{Department of Physics, Tohoku University, Sendai, Miyagi 980-8578, Japan}
\author{Hiroshi Murakawa}
\affiliation{Department of Physics, Osaka University, Toyonaka, Osaka 560-0043, Japan}
\author{Seiko Ohira-Kawamura}
\affiliation{Materials and Life Science Division, J-PARC Center, Japan Atomic Energy Agency, Tokai, Ibaraki 319-1195, Japan}
\author{Kenji Nakajima}
\affiliation{Materials and Life Science Division, J-PARC Center, Japan Atomic Energy Agency, Tokai, Ibaraki 319-1195, Japan}
\author{Hidetoshi Masuda}
\affiliation{Institute for Materials Research, Tohoku University, Sendai, Miyagi 980-8577, Japan}
\author{Masaki Fujita}
\affiliation{Institute for Materials Research, Tohoku University, Sendai, Miyagi 980-8577, Japan}
\author{Noriaki Hanasaki}
\affiliation{Department of Physics, Osaka University, Toyonaka, Osaka 560-0043, Japan}
\author{Yoshinori Onose}
\affiliation{Institute for Materials Research, Tohoku University, Sendai, Miyagi 980-8577, Japan}
\author{Yusuke Nambu}
\affiliation{Institute for Materials Research, Tohoku University, Sendai, Miyagi 980-8577, Japan}
\affiliation{Organization for Advanced Studies, Tohoku University, Sendai, Miyagi 980-8577, Japan}
\affiliation{FOREST, Japan Science and Technology Agency, Kawaguchi, Saitama 332-0012, Japan}

\date{\today}

\begin{abstract}
    Magnetic excitations in the noncentrosymmetric magnet Sr$_2$MnSi$_2$O$_7$ were investigated through inelastic neutron scattering measurements.
    Major magnetic excitations are limited up to the energy transfer of 0.5~meV, and two magnon branches under zero magnetic field were well explained in the framework of linear spin-wave theory.
    The magnitudes of the square-lattice in-plane and inter-plane nearest-neighbor interactions, spin anisotropy term, and the Dzyaloshinskii-Moriya interaction are respectively estimated to be $J_1=45.54(5)$~$\mu$eV, $J_2=0.52(1)$~$\mu$eV, $\Lambda=4.98(11)$~$\mu$eV, $D_{xy}=0.02(9)$~$\mu$eV, and $D_z=4.10(1)$~$\mu$eV, and calculations using these parameters reproduce experimental data quite well.
    Sr$_2$MnSi$_2$O$_7$ appears to have the smallest energy scale among the melilite-type compounds, and the small $J_2/J_1=0.0114(2)$ indicates the sufficient two-dimensionality.
\end{abstract}

\maketitle

\section{\label{sec:level1}Introduction}

Multiferroics, in which magnetism and dielectricity are cross-correlated, are expected to be one of the key ingredients for developing power-saving devices~\cite{Fiebig2016}.
Spin-induced multiferroics, as first identified in TbMnO$_3$~\cite{Kimura2003}, have an advantage of electrical/magnetic mutual correlation and exemplified an easy control of electric charge by magnetic signals and magnetization by electric currents~\cite{Tokura2014}.
Spin-induced dielectricity is classified into three major mechanisms: exchange striction~\cite{Hur2004}, spin current~\cite{Katsura2005}, and spin-dependent $d$-$p$ hyblidization~\cite{Arima2007}.

Melilite-type compounds are one of the representative groups with multiferroic properties in the $d$-$p$ hybridization mechanism~\cite{Murakawa2012}.
The compounds with the stoichiometry $A_2MB_2X_7$ ($A=$ Ca, Sr, Ba, $M=$ Mn, Co, Cu, $B=$ Si, Ge, $X=$ O, S)~\cite{Sazonov2016_1,Yi2008,Hutanu2011,Hutanu2012,Hutanu2014,Solovyev2015,Sazonov2016_2,Akaki2012,Akaki2017,Kurihara2024,Endo2017} have generally the $P\overline{4}2_1m$ space group and the Dzyaloshinskii-Moriya (DM) interaction inherent owing to the broken inversion symmetry.
The competition between the antiferromagnetic nearest-neighbor exchange interaction and the DM interaction naturally yields canting antiferromagnetic structures~\cite{Hutanu2012,Dutta2023} and even helical magnetic structures~\cite{Muhlbauer2012}.
The $d$-$p$ hybridization induced by $MX_4$ tetrahedra leads to a spin direction-dependent dielectricity that can easily be controlled by external magnetic fields~\cite{Murakawa2010}.
Besides that, melilite-type compounds are of broad interest for circular dichroism~\cite{Bordacs2012}, quantum fluctuations~\cite{Kim2014}, giant optical effects~\cite{Toyoda2023}, and magnon textures~\cite{Kawano2019_1}.

The melilite-type compounds possess a wide variety of magnetic structures.
The compound with $M=$ Co$^{2+}$ has the C-type antiferromagnetic structure, and the magnetic moment is projected onto the $ab$-plane~\cite{Hutanu2012,Endo2012,Dutta2023}.
On the other hand, for $M=$ Mn$^{2+}$, magnetic structures are ranged from the C-type (Ba$_2$MnSi$_2$O$_7$~\cite{Sale2019}) to the G-type (Ba$_2$MnGe$_2$O$_7$~\cite{Sazonov2018,Hasegawa2021,Sazonov2023}, Sr$_2$MnGe$_2$O$_7$~\cite{Endo2012}), which have magnetic moments either on the $ab$-plane or along the $c$-axis.
The model spin Hamiltonian is also $M$ cation dependent, where the XXZ model taking into account anisotropic exchange interaction~\cite{Soda2014,Soda2018} describes $M=$ Co$^{2+}$ ($S=3/2$) cases, whereas a Heisenberg-type exchange interaction~\cite{Masuda2010} is employed for $M=$ Mn$^{2+}$ ($S=5/2$).
To achieve a collective understanding of the magnetism of the melilite-type compounds, information on the magnetic structure and excitations over separate mixtures of $A$, $B$, and $X$ atoms is required.

\begin{figure}[t!]
    \includegraphics[width=1\linewidth]{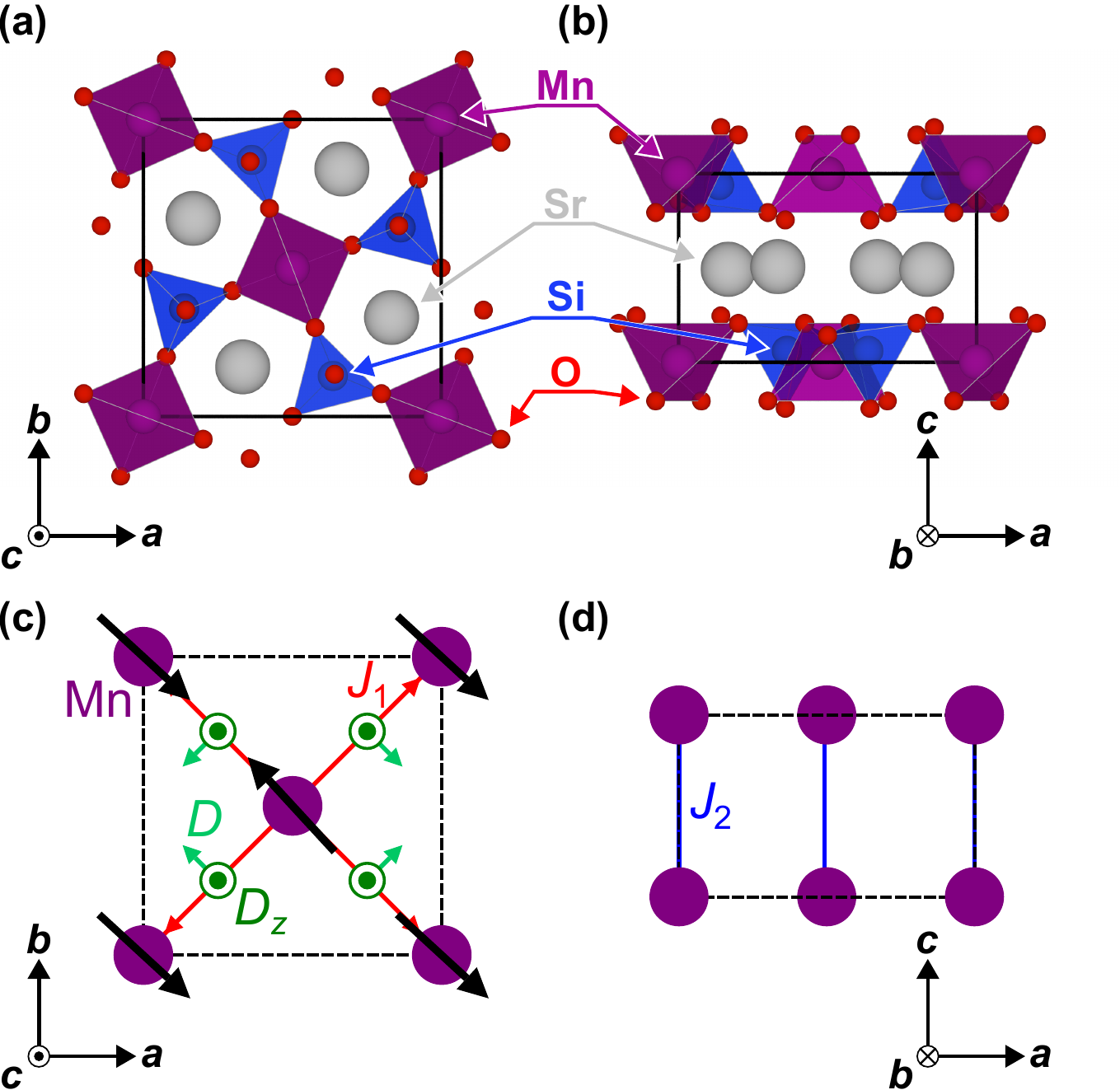}
    \caption{Crystal structure of Sr$_2$MnSi$_2$O$_7$ projected onto (a) the $ab$-plane and (b) the $ac$-plane. Assumed colinear magnetic structure and interaction pathways on (c) the $ab$ and (d) the $ac$-plane.}
    \label{f1}
\end{figure}

We here measure magnetic excitations of a new melilite-type compound, Sr$_2$MnSi$_2$O$_7$, and then determine the effective model for this particular material.
Sr$_2$MnSi$_2$O$_7$ has a two-dimensional (2D) network of MnO$_4$ and SiO$_4$ tetrahedra as illustrated in Fig.~\ref{f1}(a,b).
Mn$^{2+}$ ions with spin $S=5/2$ reside on a square lattice on the $ab$-plane~\cite{Endo2010}, which is stacked along the $c$-axis and is separated by Sr$^{2+}$ ions.
An antiferromagnetic order appears below $T_{\rm N} = 3.4$~K~\cite{Akaki2013}, and the G-type canting antiferromagnetic structure is refined~\cite{Nambu2024}.
Prior to verifying the multiferroic properties of the compound and possible magnon texture, it is crucial to construct an effective microscopic model.
In this study, we observe two magnon modes in Sr$_2$MnSi$_2$O$_7$ through an inelastic neutron scattering experiment with zero magnetic field.
The observed magnon spectra are well accounted for by the linear spin-wave theory, and the strengths of primary magnetic interactions are refined.

\section{Experimental}

Polycrystalline samples of Sr$_2$MnSi$_2$O$_7$ were synthesized by the solid-state reaction.
Single crystalline samples were then grown in the air by the floating zone method using an image furnace~\cite{Murakawa2012}.
A single crystal with a cylindrical shape was obtained, and it was cut to the size $\phi5\times30$~mm$^3$ for neutron measurements. 
The quality of the obtained samples was confirmed by X-ray diffraction, and neutron diffraction~\cite{Nambu2024} measurements.

To observe magnetic excitations in Sr$_2$MnSi$_2$O$_7$, an inelastic unpolarised neutron scattering experiment was carried out on the chopper spectrometer BL14 AMATERAS~\cite{Nakajima2011} at J-PARC, Japan.
The instrument enables effective data collection using the multiple incident energies ($E_{\rm i}$) simultaneously.
We employed $E_{\rm i}$ = 7.733, 3.136, 1.687~meV with energy resolutions of 0.273, 0.073, 0.027~meV at the elastic position, respectively.
The crystal oriented on the [$HK0$] horizontal scattering zone, was inserted into the instrument-equipped cryostat without applying magnetic fields. 

\section{Results and discussion}

\begin{figure}[t!]
    \includegraphics[width=1\linewidth]{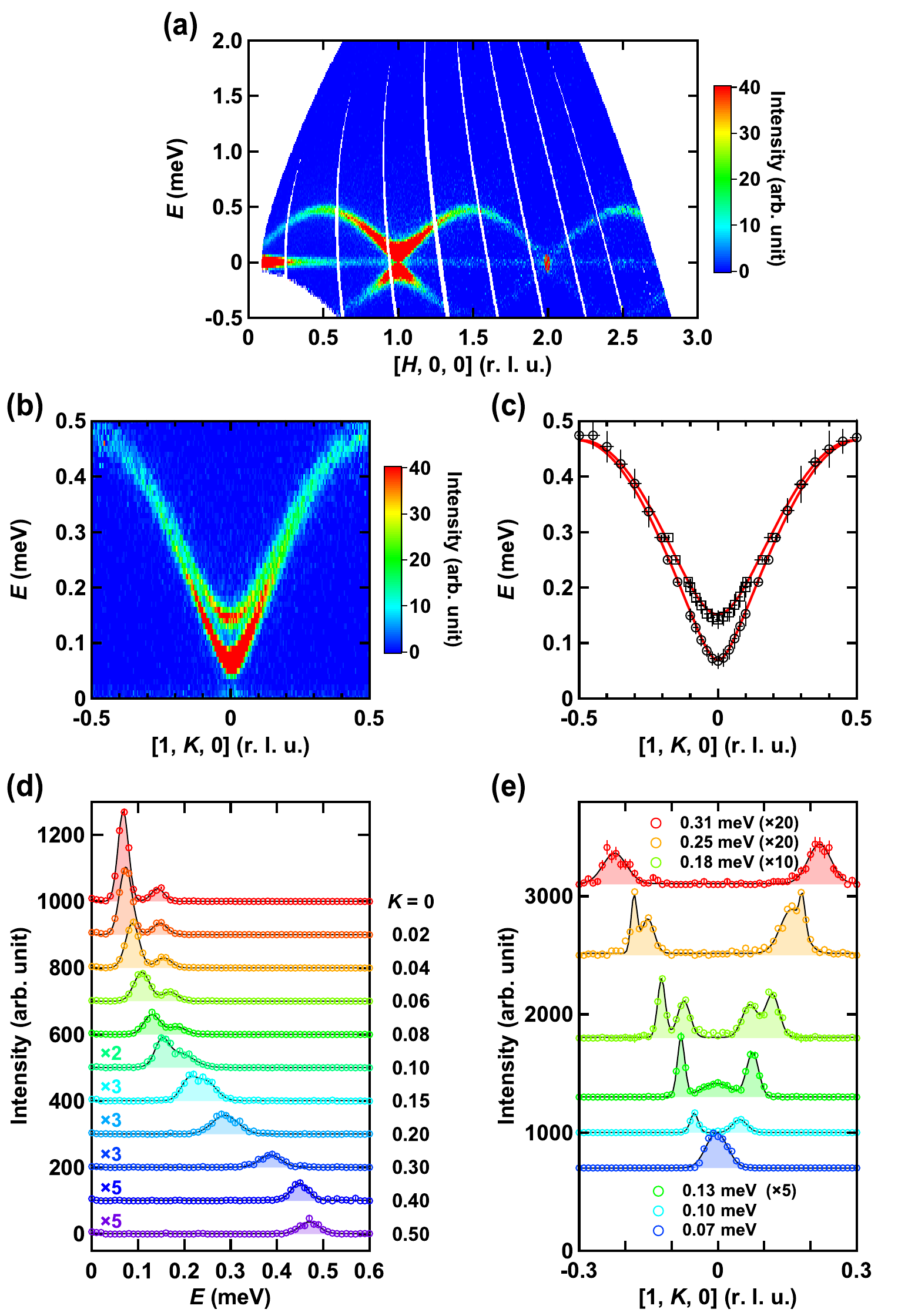}
    \caption{(a) Neutron intensity map for the [$H,0,0$]~r.l.u. direction using $E_{\rm i}=3.136$~meV and (b) enlarged map for the [$1,K,0$]~r.l.u. direction using $E_{\rm i}=1.687$~meV.  (c) Estimated peak positions of magnon spectra from (d,e), with red curves giving the fit based upon the linear spin-wave calculation. (d) One-dimensional cut along the $E$ where the $K$-integration is $\pm0.01$~r.l.u. and $0.09\le H\le 1.01$~r.l.u., together with fits using Gaussian peaks. The observed neutron-scattering intensity was multiplied for data with $K\ge 0.10$ to enlarge the observed peak. (e) One-dimensional cut along [$1,K,0$]~r.l.u. direction where the Constant $E$-integration is $\pm0.005$~meV and $0.09\le H\le 1.01$, $-0.1\le L\le 0.1$~r.l.u.}
    \label{f2}
\end{figure}

Overall magnetic excitations measured at 1.91(4)~K using $E_{\rm i}=3.136$~meV under zero magnetic field are depicted in Fig.~\ref{f2}(a), where the whole excitations reside below the energy transfer, $E=0.5$~meV.
Incoherent scattering centered at $E=0$~meV is visible yet very small reflecting the small incoherent scattering cross-sections of Sr (0.06 barn), Mn (0.4 barn), Si (0.004 barn), and O (0.0008~barn)~\cite{Sears1992}. 
The magnon dispersion using $E_{\rm i}=1.687$~meV with a tighter energy resolution is shown in Fig.~\ref{f2}(b), where two distinct branches are clearly observed.
$K$-integrated one-dimensional data along $E$ are derived and are fit by Gaussian peaks [Fig.~\ref{f2}(d)].
Two peaks are observed for smaller $K$-regime, and they merge into a single peak with increasing $K$.
The cut data along the $K$-direction [Fig.~\ref{f2}(e)] are also analyzed, and estimated peak positions are summarized in Fig.~\ref{f2}(c).

\begin{figure*}[t!]
    \includegraphics[width=1\linewidth]{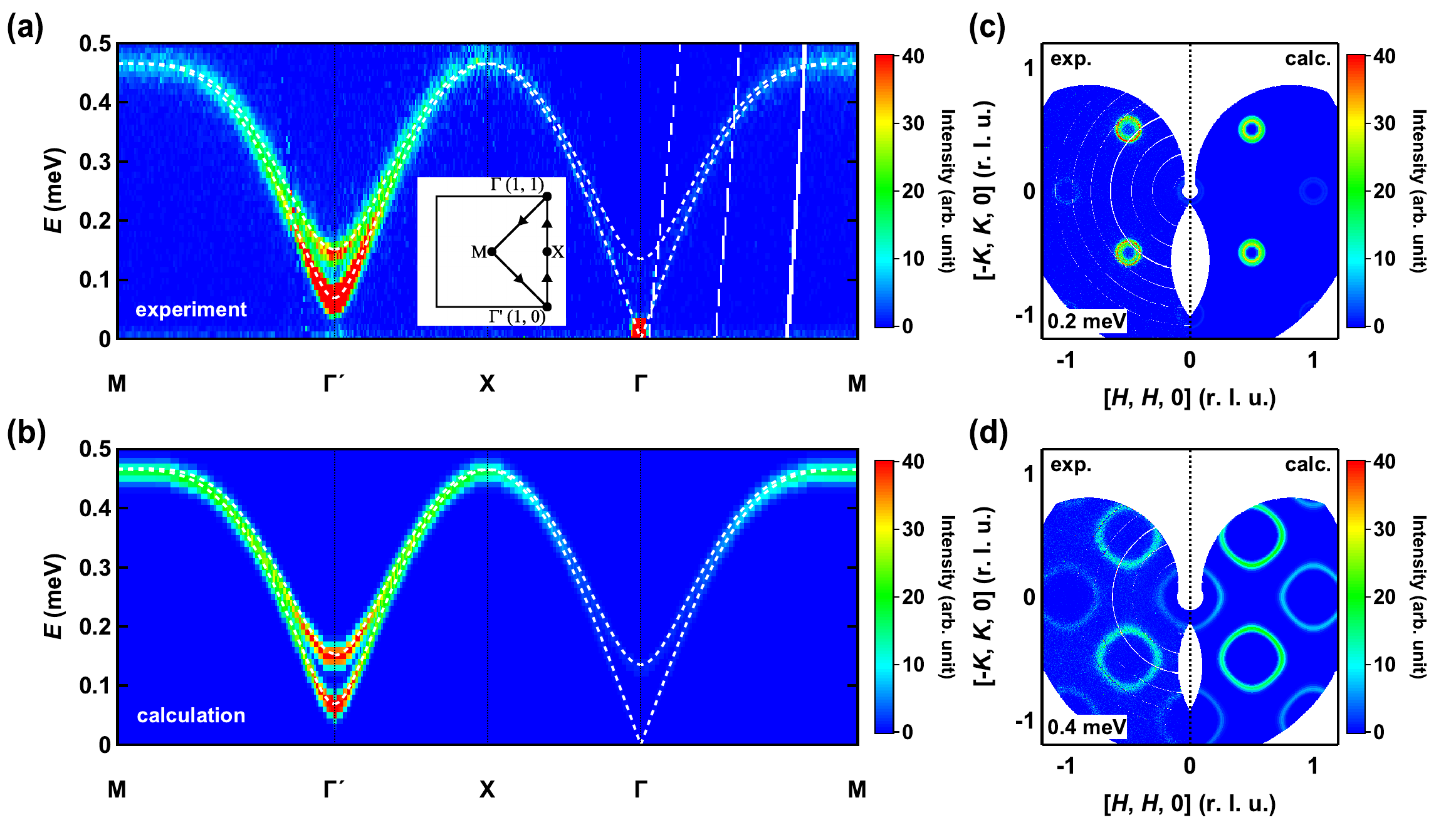}
	\caption{(a) Observed and (b) calculated neutron intensity map of Sr$_2$MnSi$_2$O$_7$ along high-symmetry directions. Measurements are performed at $T = 1.91(4)$ K ($< T_{\rm N}$), and data were integrated for $-0.1\leq L\leq 0.1$~r.l.u. and over a thickness of $\pm$0.005 r.l.u. perpendicular to the path directions. The inset to (a) draws the $[HK0]$-reciprocal space map. Constant energy slices for (c) $0.19\le E\le 0.21$~meV and (d) $0.39\le E\le 0.41$~meV are compared with the corresponding calculations.}
    \label{f3}
\end{figure*}

To evaluate exchange interactions, spin anisotropy term and the DM interaction, obtained data were analyzed within the linear spin-wave calculation.
The spin Hamiltonian we employed reads,
\begin{align}
    \mathcal{H}=&\sum_{\langle i,j\rangle}{J_1\vec{S}_i\cdot \vec{S}_j}+\sum_{\langle k,l\rangle}{J_2\vec{S}_k\cdot \vec{S}_l}\nonumber\\
    &+\sum_{i}\Lambda(\vec{S}_i^z)^2+\sum_{\langle i,j\rangle}{\vec{D}\cdot(\vec{S}_i\times\vec{S}_j)},
    \label{eq1}
\end{align}
where, $i,j,k,l$ indices stand for Mn$^{2+}$ at the Wyckoff 2$a$ site, $\langle i,j\rangle$ and $\langle k,l\rangle$ run nearest-neighbor pairs on the $ab$-plane and along the $c$-axis, respectively [corresponding pathways depicted in Fig.~\ref{f1}(c,d)].
$\Lambda$ is the spin anisotropy term, and $\Lambda>0$ corresponds to the easy-plane anisotropy.
The DM interaction $\vec{D}=D_{xy}\frac{\vec{e}_x+\vec{e}_y}{\sqrt{2}}+D_z\vec{e}_z$ is allowed reflecting the noncentrosymmetric space group (Fig.~\ref{f1}(c)).

First, assuming sufficiently small $|\vec{D}|$, we consider a collinear magnetic structure with magnetic moments along [110].
The eigenenergies ($\varepsilon_1$, $\varepsilon_2$) can analytically be derived at high-symmetry points such as $\vec{Q}^X=(1/2,0,0)$ and $\vec{Q}^{\Gamma^\prime}=(1,0,0)$~r.l.u.,
\begin{align}
    \varepsilon_1^{X}&=2S\sqrt{2(J_1+J_2)(2J_1+\Lambda)},\label{eq2}\\
    \varepsilon_2^{X}&=2S\sqrt{2J_1(2J_1+2J_2+\Lambda)},\label{eq3}\\
    \varepsilon_1^{\Gamma^\prime}&=2S\sqrt{2J_2(4J_1+\Lambda)},\\
    \varepsilon_2^{\Gamma^\prime}&=2S\sqrt{4J_1(2J_2+\Lambda)}.
\end{align}
Given $J_2\ll J_1$ in Sr$_2$MnSi$_2$O$_7$ as discussed later on, $\varepsilon_1^{X}\simeq\varepsilon_2^{X}$ holds and they can be approximated as $2S\sqrt{2J_1(2J_1+\Lambda)}$.
Through fits to the obtained data [Fig.~\ref{f2}(d)], corresponding eigenenergies are experimentally estimated as $\varepsilon_1^{X}\simeq\varepsilon_2^{X}=0.47(2)$~meV, $\varepsilon_1^{\Gamma^\prime}=0.07(1)$~meV, and $\varepsilon_2^{\Gamma^\prime}=0.14(2)$~meV.
The strengths of the magnetic interactions were then numerically obtained using the random swarm optimization method~\cite{Kennedy1995} by fit to experimental data for whole momentum transfer, $\vec{Q}$-space measured, which nicely converges and gives $J_1=45.28(5)$~$\mu$eV, $J_2=0.52(1)$~$\mu$eV, and $\Lambda=3.95(14)$~$\mu$eV.
Using these values, the eigenenergies are computed as $\varepsilon^{X}=0.4625(5)$~meV, $\varepsilon_1^{\Gamma^\prime}=0.0694(7)$~meV, and $\varepsilon_2^{\Gamma^\prime}=0.1503(2)$~meV, yielding fair agreement with analytical solutions.
These parameters can nicely reproduce the experimental data as depicted in Fig.~\ref{f2}(c).

Using the obtained parameters, we calculate the magnon spectra in the same dynamic range as measured in Fig.~\ref{f3}(a).
Figure~\ref{f3}(b) gives results that are almost completely in accordance with our experimental findings.
Magnetic excitations at constant energies are also compared in Fig.~\ref{f3}(c,d).
With increasing $E$, magnon branches spread while holding $\bar{4}$ symmetric shapes.
Such high reproducibility of experimental data validates our parameterizations.

\begin{figure}[t!]
    \includegraphics[width=1\linewidth]{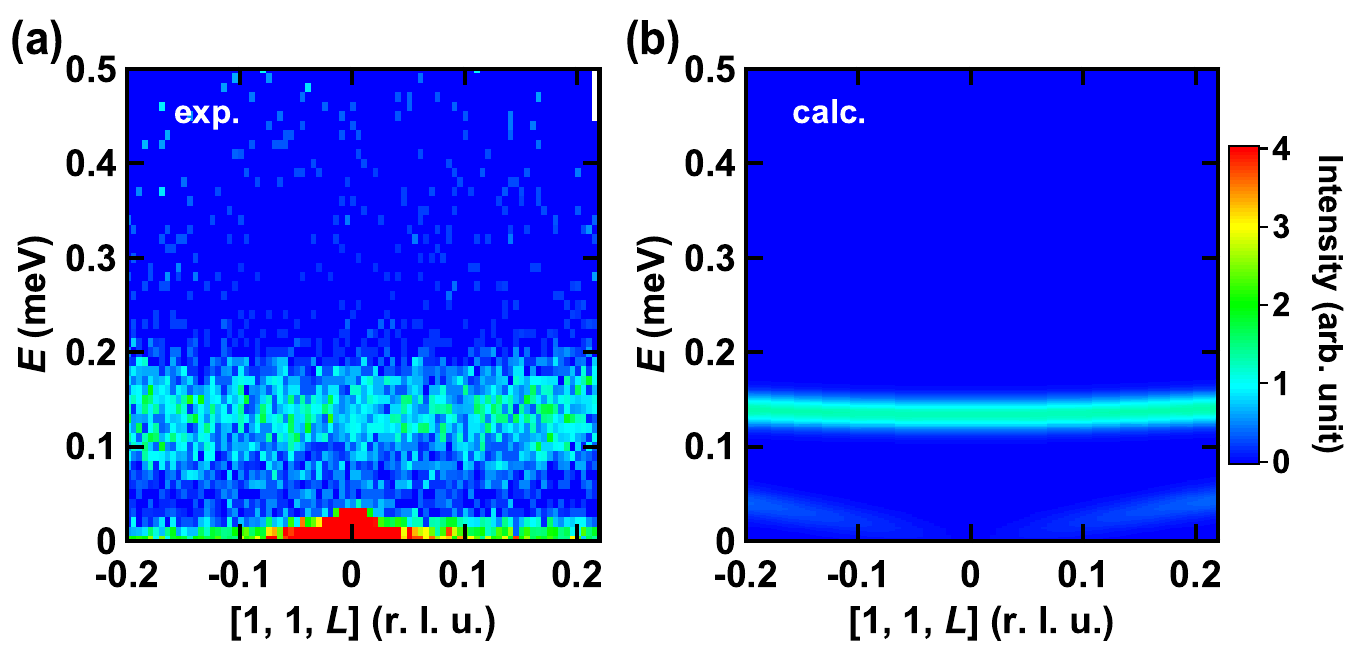}
    \caption{(a) Observed and (b) calculated neutron intensity map of Sr$_2$MnSi$_2$O$_7$ along [$1,1,L$]~r.l.u. Measurements were performed using $E_i=1.687$~meV, and data were integrated for $0.9\le H\le 1.1$~r.l.u. and $0.9\le K\le 1.1$~r.l.u.}
    \label{f4}
\end{figure}

Figure~\ref{f4}(a) shows magnon spectra along the [$1,1,L$] direction.
Since the $ab$-plane was on the horizontal scattering plane and the detector coverage along the perpendicular direction ($c$) is quite limited, data for $-0.2 \lesssim L\lesssim 0.2$~r.l.u. were only obtained.
In Fig.~\ref{f4}(a), allowed nuclear reflection at (1,1,0) is visible, and a weak magnetic signal was observed between 0.1 and 0.2~meV.
Our calculation [Fig.~\ref{f4}(b)] can roughly reproduce the experimental data including the dispersionless magnon branch reflecting the small $J_2$.
However, the experimental data clearly have a rather broad spread along $E$ than the calculation.
This may be due to unmeeting the decent scattering condition and/or capturing some peculiar dynamics, and further measurements on $[H0L]$ or $[HHL]$ zones are thus future perspectives.

Among the above-mentioned refined parameters, $J_1>0$ and $J_2>0$ stand for antiferromagnetic interactions for the intra- and inter-planes.
This stabilizes the G-type antiferromagnetic structure being the same as in Sr$_2$MnGe$_2$O$_7$ and Ba$_2$MnGe$_2$O$_7$~\cite{Endo2012,Masuda2010}.
The magnetic interactions of Sr$_2$MnSi$_2$O$_7$ are slightly smaller than Ba$_2$MnGe$_2$O$_7$ with $J_1=55.6(6)$~$\mu$eV and $J_2=2.0(2)$~$\mu$eV~\cite{Masuda2010}.
The values of $J_2/J_1$ are 0.0114(2) and 0.036(4) for Sr$_2$MnSi$_2$O$_7$ and Ba$_2$MnGe$_2$O$_7$, respectively.
The smaller $J_2/J_1$ indicates that Sr$_2$MnSi$_2$O$_7$ has much sufficient two-dimensionality than Ba$_2$MnGe$_2$O$_7$.

We now move on to the spin anisotropy term $\Lambda$.
In $M=$ Co$^{2+}$ systems among the melilite-type compounds, unquenched orbital degree of freedom is evidenced by the temperature-dependent susceptibility~\cite{Hutanu2014}, leading to three orders of magnitude larger spin anisotropy, such as $\Lambda=1.034$~meV for Ba$_2$CoGe$_2$O$_7$~\cite{Soda2014}.
On the other hand, in $M=$ Mn$^{2+}$ systems, $\Lambda$ stays in $\mu$eV orders reflecting the isotropic spin as $\Lambda=2.06$~$\mu$eV for Ba$_2$MnGe$_2$O$_7$~\cite{Hasegawa2021}.
The weak easy-plane anisotropy is owing to magnetic dipole-dipole interaction~\cite{Koo2012}, and in fact, the magnetization process behaves weakly anistropic~\cite{Akaki2013}.

\begin{figure}[t!]
    \includegraphics[width=1\linewidth]{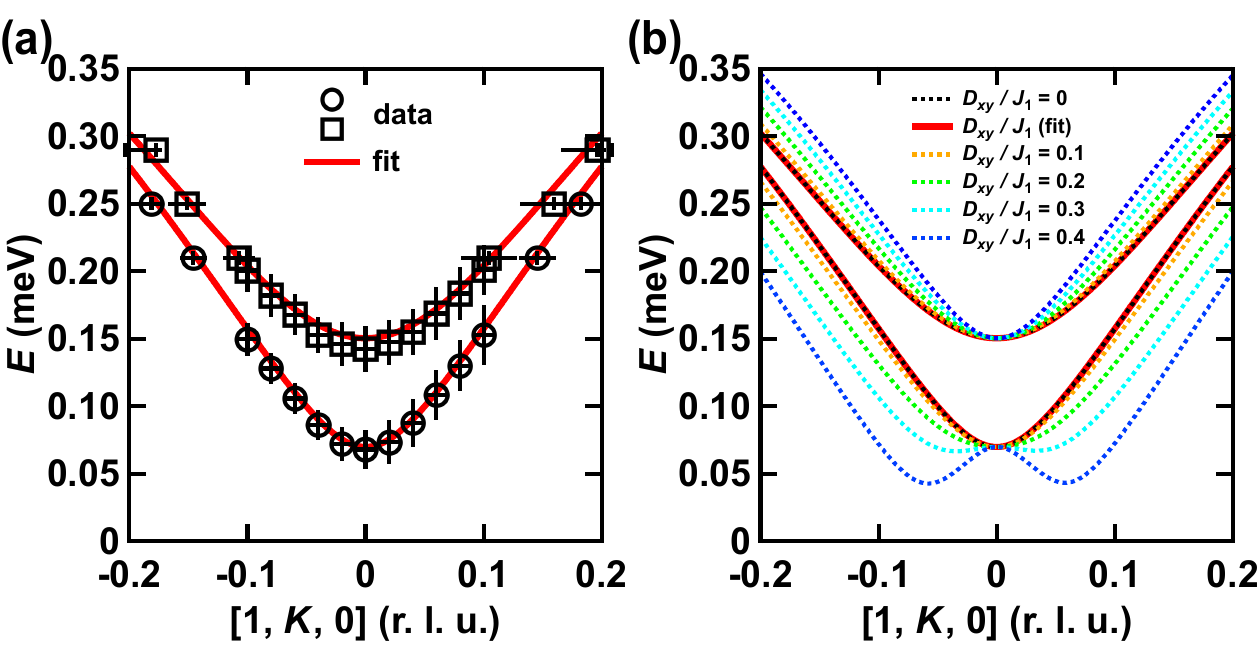}
    \caption{(a) Estimated peak positions which are identical to Fig.~\ref{f2}(c) and fit results using eq.~(\ref{eq1}). (b) Simulated magnon branches for changing $D_{xy}/J_1$ from 0.0 to 0.4 (dashed curves), where fit results give the solid red curve ($D_{xy}=0.02(9)$; $D_{xy}/J_1=0.4(2.0)\times10^{-4}$).}
    \label{f5}
\end{figure}

Next, by turning $\vec{D}$ on, the magnetic structure is no longer of collinear.
The $c$-component of the DM interaction, $D_z$, competes with $J_1$ and then stabilizes a canted antiferromagnetic structure.
With nearest-neighbor spins $\vec{S}_1$, $\vec{S}_2$ and their interactions $J_1$, $D_z$, the internal energy is written as follows,
\begin{align}
    E=J_1\vec{S}_1\cdot\vec{S}_2+D_z(\vec{S}_1\times\vec{S}_2)_z.
    \label{eq6}
\end{align}
By minimizing this energy, a relationship between $D_z/J_1$ and $\theta$ being the canting angle can be derived as $|\theta|=\frac{1}{2}\tan^{-1}{D_z/J_1}$.
Recent magnetic structure study has refined that Sr$_2$MnSi$_2$O$_7$ has $D_z/J_1=0.09$ with $\theta\sim2.4^\circ$ from powder diffraction data~\cite{Nambu2024}.
Given that the estimation was made via the spherically $\vec{Q}$-integration and four sorts of domain formations exist, $D_z/J_1=0.09$ thus poses the lower minimum.
The $ab$-component of the DM interaction, $D_{xy}$, makes off-diagonal components finite in the spin Hamiltonian, affecting the magnon dispersion shapes at reciprocal places away from the $\Gamma$ points.
Using the fixed $D_z/J_1=0.09$ and variable $D_{xy}$, further numerical fit was performed, and obtained results are $J_1=45.54(5)$~$\mu$eV, $J_2=0.52(1)$~$\mu$eV, $\Lambda=4.98(11)$~$\mu$eV, $D_{xy}=0.02(9)$~$\mu$eV, $D_z=4.10(1)$~$\mu$eV (Fig.~\ref{f5}(a)).
As simulated in Fig.~\ref{f5}(b), with increasing $D_{xy}$, the lower-energy magnon dispersion should have two local minima~\cite{Kawano2019_2}.
Note that the evaluated $D_{xy}$ under zero field appears to be negligibly small reflecting the presence of the four sorts of magnetic domains averaging out the asymmetry of the dispersion.
Our analysis on the peak spread along $\vec{Q}$ at (1,0,0) poses the upper maximum, $D_{xy}/J_1\lesssim 0.28$.
Experiments under finite fields will be needed to precisely determine $D_{xy}$.

We here compare the inplane nearest-neighbor exchange interaction $J_1$ and the whole excitation energies.
Again Sr$_2$MnSi$_2$O$_7$ has $J_1=45.5$~$\mu$eV, whereas $J_1=55.6$~$\mu$eV~\cite{Masuda2010,Thoma2024} for Ba$_2$MnGe$_2$O$_7$, $J_1=208$~$\mu$eV~\cite{Soda2014,Hutanu2014} for Ba$_2$CoGe$_2$O$_7$, and $J_1=150$~$\mu$eV~\cite{Jang2021} for Ba$_2$FeSi$_2$O$_7$.
Interestingly, $M=$ Mn$^{2+}$ systems with the largest $S$ have smaller $J_1$ values than the Co$^{2+}$ and Fe$^{2+}$ cases.
As eq.~\ref{eq2} and \ref{eq3} formulate, the whole energy scale, i.e., the upper maximum at the zone boundary ($\vec{Q}^X$), is $S$-size dependent.
Experimentally determined upper maximum in Sr$_2$MnSi$_2$O$_7$ is 0.5~meV, which is comparable with 0.6~meV for Ba$_2$MnGe$_2$O$_7$~\cite{Masuda2010} but is quite smaller than 2~meV for Ba$_2$CoGe$_2$O$_7$~\cite{Soda2014} and 2.5~meV for Ba$_2$FeSi$_2$O$_7$~\cite{Do2023}.
The fact is owing to the small $J_2$ and $\Lambda$ in Sr$_2$MnSi$_2$O$_7$, and the target compound turns out to have the smallest energy scale among the melilite-type compounds with known strengths of the magnetic interactions.

To more closely examine magnetic interactions compared with the sibling Ba$_2$MnGe$_2$O$_7$, super-exchange pathways are examined.
In the melilite-type compounds, super-exchange pathways for $J_1$ and $J_2$ require electron hopping via two oxygens like Mn-O-O-Mn, and we use the structural parameters refined at 10~K~\cite{Nambu2024,Sazonov2018}.
For $J_1$ within the plane, bond angles are 141.34(9)$^\circ$ for Sr$_2$MnSi$_2$O$_7$ and 141.73(3)$^\circ$ for  Ba$_2$MnGe$_2$O$_7$.
For $J_2$ in inter-plane, on the other hand, bond angles are 92.40(7)$^\circ$ for Sr$_2$MnSi$_2$O$_7$ and 93.97(2)$^\circ$ for Ba$_2$MnGe$_2$O$_7$.
The differences in these angles fall within a few degrees, yet yielding more than 18\% differences in $J_1$ and $J_2$.Besides that, the Kanamori-Goodenough rule~\cite{Kanamori1959,Goodenough1955} predicts that interactions are antiferromagnetic and ferromagnetic when the bond angles are close enough to 180$^\circ$ and 90$^\circ$, respectively.
However, this is not the case for both compounds given both $J_1$ and $J_2$ are antiferromagnetic.
The difference in the magnetic interactions may be due to variations of the wave function overlaps reflecting separate ionic radii.
The first-principle calculations succeeded in qualitatively explaining the difference of the interactions between Sr$_2$MnSi$_2$O$_7$ and Ba$_2$MnGe$_2$O$_7$~\cite{Koo2012}.
Our refined parameters of magnetic interactions would help enhance the accuracy of calculations.

\section{Conclusions}

To summarize, we observed the magnon spectra of the noncentrosymmetric antiferromagnet Sr$_2$MnSi$_2$O$_7$ through the inelastic neutron scattering experiment on a single-crystalline sample.
Through the analyses of magnetic excitations up to 0.5~meV, the primary magnetic interactions are successfully refined based upon the linear spin-wave theory, and the experimental data are actually well reproduced using such parameters.
The target compound stays at the smallest energy scale among the melilite-type compounds, and the tiny $J_2/J_1$ implies the sufficient two-dimensionality inherent.

\begin{acknowledgments}
We thank T. Arima, H. Kawamura, and T. Koretsune, J. Nasu for their valuable discussions.
This work was supported by the JSPS (Nos.~22KJ0312, 21H03732, 22H05145, 24K00572, 24H01638, 24H00189), FOREST (No. JPMJFR202V) and SPRING from JST, and the Graduate Program in Spintronics at Tohoku University.
Work at J-PARC was performed under the user program (No.~2020B0105).
\end{acknowledgments}

\bibliographystyle{}

\end{document}